\documentclass[runningheads]{llncs}
\usepackage[misc]{ifsym}
\usepackage{epstopdf}
\usepackage{graphicx}
\usepackage{booktabs}
\usepackage{multirow}
\begin{document}
\title{Residual Spatial Attention Network for Retinal Vessel Segmentation\thanks{This work is supported by the China Scholarship Council, the Stipendium Hungaricum Scholarship, the National Natural Science Foundation of China under Grants 61602221 and 61672150, the Chinese Postdoctoral Science Foundation 2019M661117, and Hungarian Government and co-financed by the European Social Fund (EFOP-3.6.3-VEKOP-16-2017-00001).}}
%
%

\author{Changlu Guo\inst{1, 2} $^{(\textrm{\Letter})}$ \and
M{\'a}rton Szemenyei\inst{2} \and
Yugen Yi\inst{3} $^{(\textrm{\Letter})}$ \and
Wei Zhou\inst{4}\and
Haodong Bian\inst{5}
}
\authorrunning{Guo et al.}

\institute{E{\"o}tv{\"o}s Lor{\'a}nd University, Budapest, Hungary\\
\email{clguo.ai@gmail.com}\\
\url{https://github.com/clguo/RSAN}\\
\and
Budapest University of Technology and Economics, Budapest, Hungary \\
\and
Jiangxi Normal University, Nanchang, China\\
\email{yiyg510@jxnu.edu.cn}\\
\and
Chinese Academy of Science, Shengyang, China\\
\and
Qinghai University, Xining, China\\
}
\maketitle              
\begin{abstract}
Reliable segmentation of retinal vessels can be employed as a way of monitoring and diagnosing certain diseases, such as diabetes and hypertension, as they affect the retinal vascular structure. In this work, we propose the Residual Spatial Attention Network (RSAN) for retinal vessel segmentation. RSAN employs a modified residual block structure that integrates DropBlock, which can not only be utilized to construct deep networks to extract more complex vascular features, but can also effectively alleviate the overfitting. Moreover, in order to further improve the representation capability of the network, based on this modified residual block, we introduce the spatial attention (SA) and propose the Residual Spatial Attention Block (RSAB) to build RSAN. We adopt the public DRIVE and CHASE DB1 color fundus image datasets to evaluate the proposed RSAN. Experiments show that the modified residual structure and the spatial attention are effective in this work, and our proposed RSAN achieves the state-of-the-art performance.

\end{abstract}

\section{Introduction}
Retinal images contain rich contextual structures, such as retinal vascular structures that can provide important clinical information for the diagnosis of diseases such as diabetes and hypertension. Therefore, the accuracy of retinal blood vessel segmentation can be used as an important indicator for the diagnosis of related diseases. However, manual segmentation of retinal blood vessels is a time-consuming task, so we are working to find a way to automatically segment retinal blood vessels.

In recent years, convolutional neural network (CNN) based methods have shown strong performance in automatically segmenting retinal blood vessels. In particular, Ronneberger et al. \cite{1} proposed the famous U-Net that combines coarse features with fine features through ``skip connections" to have superior performance in the field of medical image processing. Therefore, numerous retinal vessel segmentation methods are based on U-Net, for example, Zhuang et al. \cite{2} proposed a chain of multiple U-Nets (LadderNet), which includes multiple pairs of encoder-decoder branches. Wu et al. \cite{3} reported the multi-scale network followed network (MS-NFN) for retinal vessel segmentation and each sub-model contains two identical U-Net models. Then Wang et al. \cite{4} proposed the Dual Encoding U-Net (DEU-Net) that remarkably enhances networks capability of segmenting retinal vessels in an end-to-end and pixel-to-pixel way. Although these U-Net-based methods have achieved excellent performance, they all ignore the inter-spatial relationship between features, which is important for retinal vessel segmentation because the distinction between vascular and non-vascular regions in the retinal fundus image is not very obvious, especially for thin blood vessels. To address this problem, we introduce spatial attention (SA) in this work because it can learn where is able to effectively emphasize or suppress and refine intermediate features \cite{5} .

In this paper, we propose a new Residual Spatial Attention Network (RSAN) for segmentation of retinal vessels in retinal fundus images. Specifically, inspired by the success of DropBlock \cite{6} and residual network \cite{7}, we add DropBlock to the pre-activation residual block \cite{8}, which can be used to build a deep network to obtain deeper vascular features. Then, based on the previous discussion, we integrate the SA into this modified residual block and propose a Residual Spatial Attention Block (RSAB). Finally, combined with the advantage that ``skip connection" in U-Net is able to save more structural information, the original convolution unit of U-Net is replaced by the modified residual block and RSAB to form the proposed RSAN. Through comparison experiments, we observe that the use of DropBlock can improve the performance. Then, after the introduction of SA, that is, using the proposed RSAN for retinal vessel segmentation, our performance surpasses other existing state-of-the-art methods.

\section{Related work}

\subsection{U-Net}
In 2015, Ronneberger et al. \cite{1} proposed a U-shaped fully convolutional network for medical image segmentation called U-Net, which has a typical symmetrical codec network structure. U-Net has an obvious advantage that it can make good use of GPU memory. This advantage is mainly related to extraction of image features at multiple image scales. U-Net transfers the feature maps obtained in the encoding stage to the corresponding decoding stage, and merges the feature maps of different stages through ``skip connection" to merge coarse and fine-level dense predictions.

\subsection{ResNet}
He et al. \cite{7} observed that when deeper networks begin to converge, there will be a degradation problem: as the network deepens, the accuracy quickly degrades after reaching saturation. In other words, simply deepening the network can hinder training. To overcome these problems, the residual network proposed by He et al. shows significantly improved training characteristics, allowing the network depth to be previously unachievable. The residual network consists of some stacked residual blocks, and each residual block can be illustrated as a routine form:
\begin{equation}
\begin{array}{l}
 {y_i} = F({x_i},{w_i}) + h({x_i}) \\
 {x_{i + 1}} = \sigma ({y_i}) \\
\end{array}
\end{equation}
where ${x_i}$ and ${x_j}$ represent the input and output of the current residual block, $\sigma ({y_i}) $
is an activation function, $F(\bullet)$ is the residual function, and $h({x_i})$ is an identity mapping function, typically
$h({x_i}) = {x_i}$.

\begin{figure}
\includegraphics[width=1\textwidth]{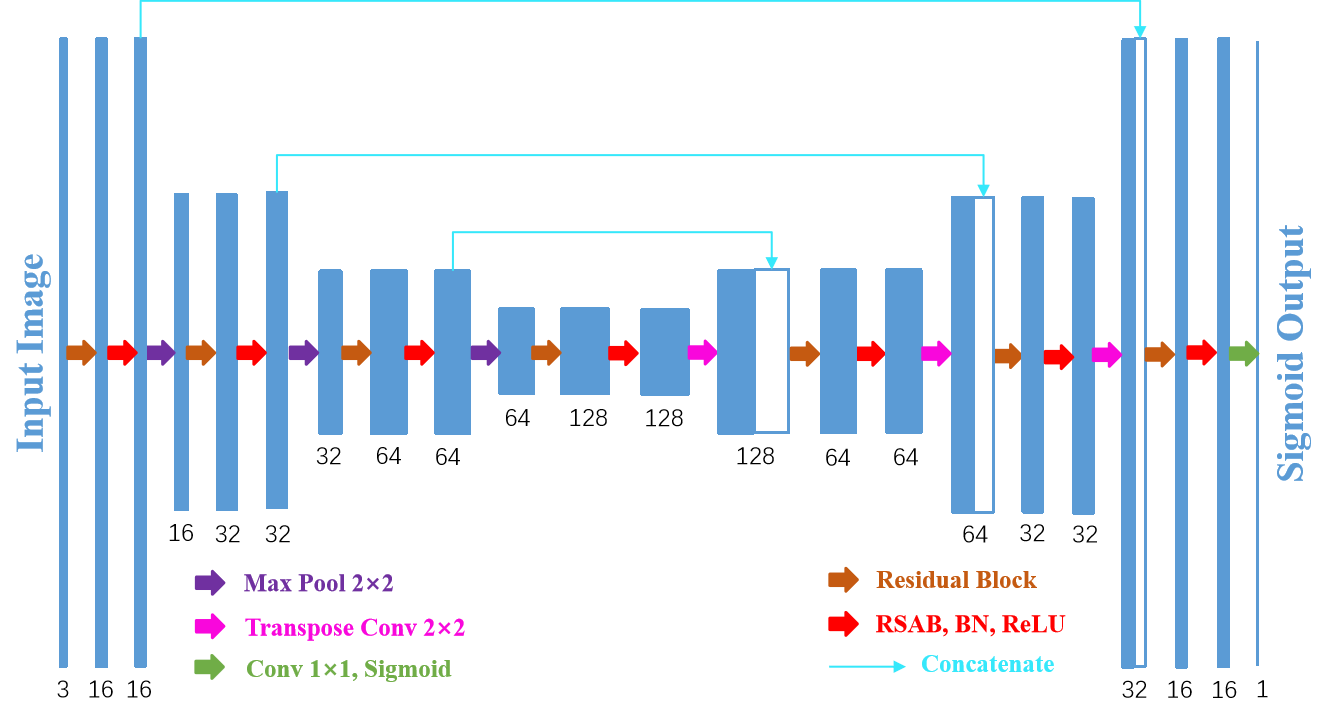}
\caption{Diagram of the proposed RSAN.} \label{fig1}
\end{figure}

\section{Method}
Figure 1 illustrates the proposed Residual Spatial Attention Network (RSAN) with a typical encoder-decoder architecture. RSAN consists of three encoder blocks (left side) and three decoder blocks (right side) that are connected by a concatenate operation. Each encoder block and decoder block contain a pre-activation residual block with DropBlock, a Residual Spatial Attention Block (RSAB), a Batch Normalization (BN) layer, and a Rectified Linear Unit (ReLU). In the encoder, the max pooling with a pooling size of 2 is utilized for downsampling, so that the size of the image after each RSAB is halved, which is beneficial to reduce the computational complexity and save training time. The decoder block and the encoder block is similar, except that the former uses a $2\times2$ transposed convolution for upsampling instead of the pooling layer. The last layer utilizes a $1\times1$ convolution followed by a Sigmoid activation function to obtain the required feature map.

\begin{figure}
\includegraphics[width=1\textwidth]{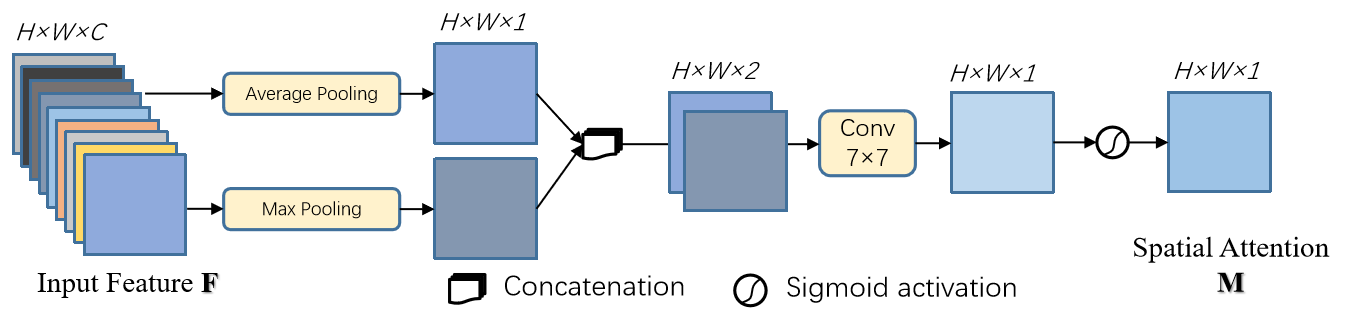}
\caption{Diagram of the Spatial Attention.} \label{fig2}
\end{figure}

\subsection{Spatial Attention}
Spatial Attention (SA) was introduced as a part of the convolutional block attention module for classification and detection \cite{5} . SA employs the inter-spatial relationship between features to produce a spatial attention map, as shown in Fig. 2. The spatial attention map enables the network to enhance important features (e.g. vascular features) and suppress unimportant ones. To obtain the spatial attention map, different from the $1\times1$ convolution commonly used in past work, SA first applies max-pooling and average-pooling operations along the channel axis and concatenates them to produce an efficient feature descriptor. The reason behind this is that the amount of SA parameters will be very small. A single SA module contains only 98 parameters, but it can bring significant performance improvements. Generally, the input feature $F\in {R^{H \times W \times C}}$ through the channel-wise max-pooling and average-poling generate ${F_{mp}} \in {R^{H \times W \times 1}}$ and ${F_{ap}} \in {R^{H \times W \times 1}}$, respectively, e.g., at the ${i}$-th pixel in ${F_{mp}}$ and $F_{ap}$:
\begin{equation}
F_{mp}^i = Max({P^{(i,c)}}),0 < c < C,0 < i < H \times W
\end{equation}
\begin{equation}
F_{ap}^i = \frac{1}{C}\sum\limits_{c = 1}^C {} ({P^{(i,c)}}),0 < c < C,0 < i < H \times W
\end{equation}
where $Max(\cdot)$ obtain the maximum number, ${P^{(i,c)}}$ represents the pixel value of the ${i}$-th pixel at the ${c}$-th channel, and ${H}$, ${W}$, and ${C}$ denote the height, width, and the number of channels for the input feature ${F}$, respectively. Then a convolutional layer followed by a Sigmoid activation function on the concatenated feature descriptor which is utilized to produce a spatial attention map $M(F)\in{R^{H\times W\times 1}}$. Briefly, the spatial attention map is calculated as:
\begin{equation}
M(F) = \sigma ({f^{7 \times 7}}([{F_{mp}};{F_{ap}}]))
\end{equation}
where ${f^{7 \times 7}}(\cdot)$ denotes a convolution operation with a kernel size of 7 and $\sigma (\cdot)$ represents the Sigmoid function.

\subsection{Modified Residual Block}
In this work, shallow networks may limit the network's ability to extract the vascular features required \cite{9}. We argue that building deeper neural networks can learn more complex vascular features, but He et al. \cite{7} observed that simply increasing the number of network layers may hinder training, and degradation problems may occur. In order to address the above problems, He et al. \cite{7} proposed the residual network (ResNet) achieving a wide influence in the field of computer vision. Furthermore, He et al. \cite{8} discussed in detail the effects of the residual block consisting of multiple combinations of ReLU activation, Batch normalization (BN), and convolutional layers, and proposed a pre-activation residual block, as shown in Fig. 3(b). We utilize this pre-activation residual block to replace the original convolutional unit of U-Net shown in Fig. 3(a), and call this modified network as``Backbone".

In addition, Ghiasi et al. \cite{6} proposed DropBlock, a structured variant of dropout, and also proved its effectiveness in convolutional networks, then SD-Unet \cite{10} and DRNet \cite{11} showed that DropBlock can effectively prevent overfitting problems in fully convolutional networks (FCNs). Inspired by the above work, we introduce DropBlock in the pre-activation residual block, as shown in Fig. 3(c). If the numbers of input and output channels are different, we employ $1\times 1$ convolution to compress or expand the number of channels.
\begin{figure}
\includegraphics[width=1\textwidth]{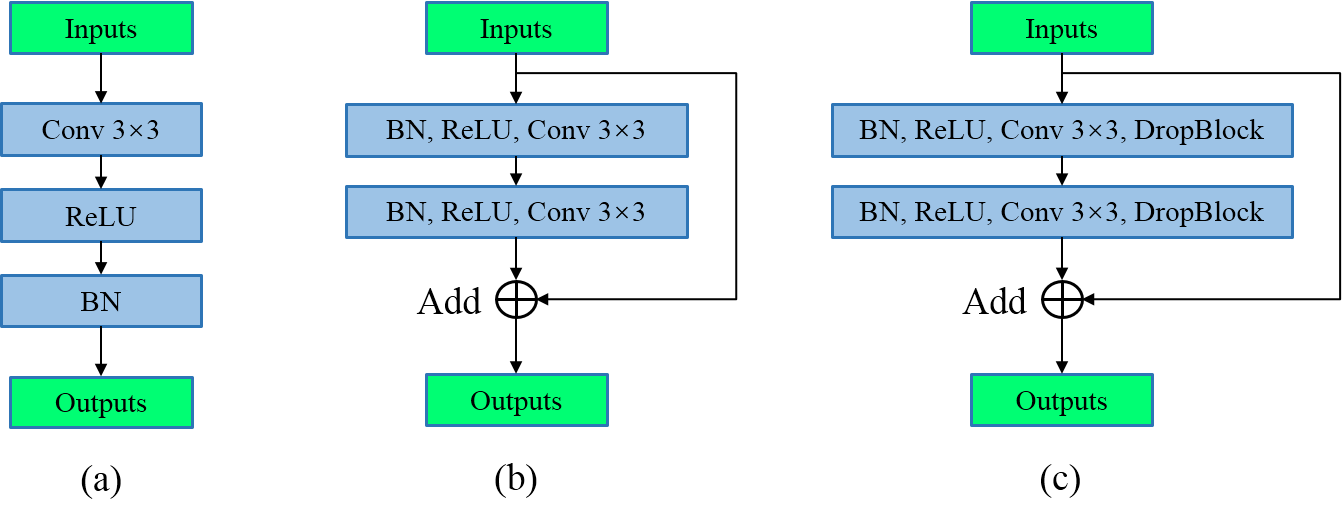}
\caption{(a) Convolutional unit of U-Net, (b) pre-activation residual block, (c) pre-activation residual block with DropBlock.} \label{fig3}
\end{figure}

\subsection{Residual Spatial Attention Block}
Spatial Attention automatically learns the importance of each feature spatial through learning, and uses the obtained importance to emphasize features or suppress features that are not important to the current retinal vessel segmentation task. Combining the previous  discussion and inspired by the successful application of the convolutional block attention module in classification and detection, we integrate SA into the modified residual block shown in Fig. 3(c) and propose the Residual Spatial Attention Block (RSAB). The structure of RSAB is shown in Fig. 4, and we argue that the introduction of SA can make full use of the inter-spatial relationship between features to improve the network's representation capability, and moreover, the integration of DropBlock and pre-activation residual block is effective without worrying about overfitting or degradation problems, even for small sample datasets such as retinal fundus image datasets.
\begin{figure}
\includegraphics[width=1\textwidth]{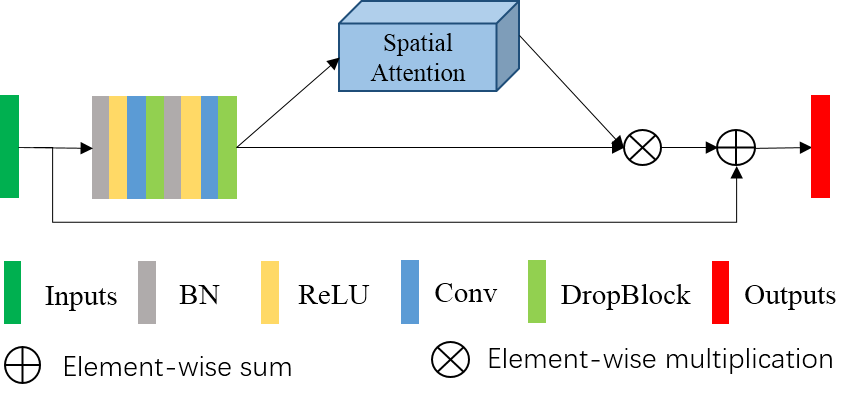}
\caption{Diagram of the proposed RSAB.} \label{fig4}
\end{figure}

\section{Experiments}

\subsection{Datasets}
We employ DRIVE and CHASE DB1 retinal image datasets to evaluate the proposed RSAN. The DRIVE dataset includes 40 color fundus images, of which 20 are officially designated for training and 20 for testing. The CHASE DB1 contains 28 retinal fundus images. Although there is no initial division of training and testing sets, usually the first 20 are used as the training set, and the remaining 8 are used as the testing set \cite{3,12} . The resolution of each image in DRIVE and CHASE DB1 is $565\times584$ and $999\times960$ respectively. In order to fit our network, we resize each image in DRIVE and CHASE DB1 to $592\times592$ and $1008\times1008$ by padding it with zero in four mar-gins, but in the process of evaluating, we crop the segmentation results to the initial resolution. The manual segmented binary vessel maps of both datasets provided by human experts can be applied as the ground truth.

\subsection{Evaluation Metrics}
To evaluate our model quantitatively, we compare the segmentation results with the corresponding ground truth and divide the comparison results of each pixel into true positive (TP), false positive (FP), false negative (FN), and true negative (TN). Then, the sensitivity (SEN), specificity (SPE), F1-score (F1), and accuracy (ACC) are adopted to evaluate the performance of the model. To further evaluate the performance of our model, we also utilize the Area Under the ROC Curve (AUC). If the value of AUC is 1, it means perfect segmentation.

\subsection{Implementation details}
We compare the performance of Backbone, Backbone+DropBlock and the proposed RSAN on DRIVE and CHASE DB1. All three models are trained from scratch using the training sets and tested in the testing sets. In order to observe whether the current training model is overfitting, we randomly select two images as the validation set from the training set of both datasets. For both datasets, we utilize the Adam optimizer to optimize all models with binary cross entropy as the loss function. During the training of DRIVE, we set the batch size to 2. RSAN first trains 150 epochs with the learning rate of $1\times{10^{-3}}$, and the last 50 epochs with $1\times{10^{-4}}$. For CHASE DB1, the batch size is 1, and a total of 150 epochs are trained, of which the first 100 epochs with a learning rate of $1\times{10^{-3}}$ and the last 50 epochs with $1\times{10^{-4}}$.

For the setting of DropBlock in RSAN, the size of block to be dropped for all datasets is set to 7. To reach the best performance, we set the probability of keeping a unit for DRIVE and CHASE DB1 to 0.85 and 0.78, respectively. In the experiments, Backbone+DropBlock and RSAN have the same configuration. For Backbone, We observed serious overfitting problems, so we use the results obtained from its best training epochs.

\subsection{Results}

Figure 5 displays some examples of two color fundus images from the DRIVE and CHASE DB1 datasets, segmentation results performed by Backbone, Backbone+DropBlock and  RSAN, and the corresponding ground truth. The segmentation results illustrate that RSAN can predict most thick and thin blood vessels (pointed by red and green arrows) comparing with Backbone and Backbone+DropBlock. In particular, RSAN has a clearer segmentation result for thin blood vessels, and can retain more detailed vascular space structure.
In addition, we quantitatively compare the performance of Backbone, Backbone+DropBlock and the proposed RSAN on the DRIVE, CHASE DB1 datasets, as displayed in Tables 1 and 2. From the results in these table, we can get several notable observations: First, Backbone+DropBlock has better performance than the Backbone, which shows that the strategy of using the DropBlock to regularize the network is effective. Second, the SEN, F1, ACC, and AUC of RSAN on the two datasets are higher than Backbone+DropBlock about 2.41 ${\%}$/1.56${\%}$, 1.12${\%}$/1.21${\%}$, 0.14${\%}$/0.15${\%}$, and 0.33${\%}$/0.23${\%}$, respectively. It proves that the introduction of spatial attention can improve the performance of the network in retinal vessel segmentation task. At last, our proposed RSAN has the best segmentation performance overall, which means that RSAN is an effective method for retinal vessel segmentation.

\begin{table}[htbp]
  \centering
  \caption{Experimental results on DRIVE.(*The results is obtain from \cite{14})}
    \begin{tabular*}{\hsize}{@{}@{\extracolsep{\fill}}cccccc@{}}
    \toprule
     Datasets & \multicolumn{5}{c}{DRIVE} \\
    \midrule
    Metrics & SEN & SPE & F1 & ACC & AUC \\
    \midrule
    U-Net \cite{5}* & 0.7537 & 0.9820 & 0.8142 & 0.9531 & 0.9755 \\
    Backbone & 0.7851 & 0.9826 & 0.7985 & 0.9653 & 0.9762 \\
    Backbone+DropBlock & 0.7908 & \textbf{0.9847} & 0.8110 & 0.9677 & 0.9822 \\
    \textbf{RSAN} & \textbf{0.8149} & 0.9839 & \textbf{0.8222} & \textbf{0.9691} & \textbf{0.9855} \\
    \bottomrule
    \end{tabular*}%
  \label{tab:addlabel}%
\end{table}%

\begin{figure}
\includegraphics[width=1\textwidth]{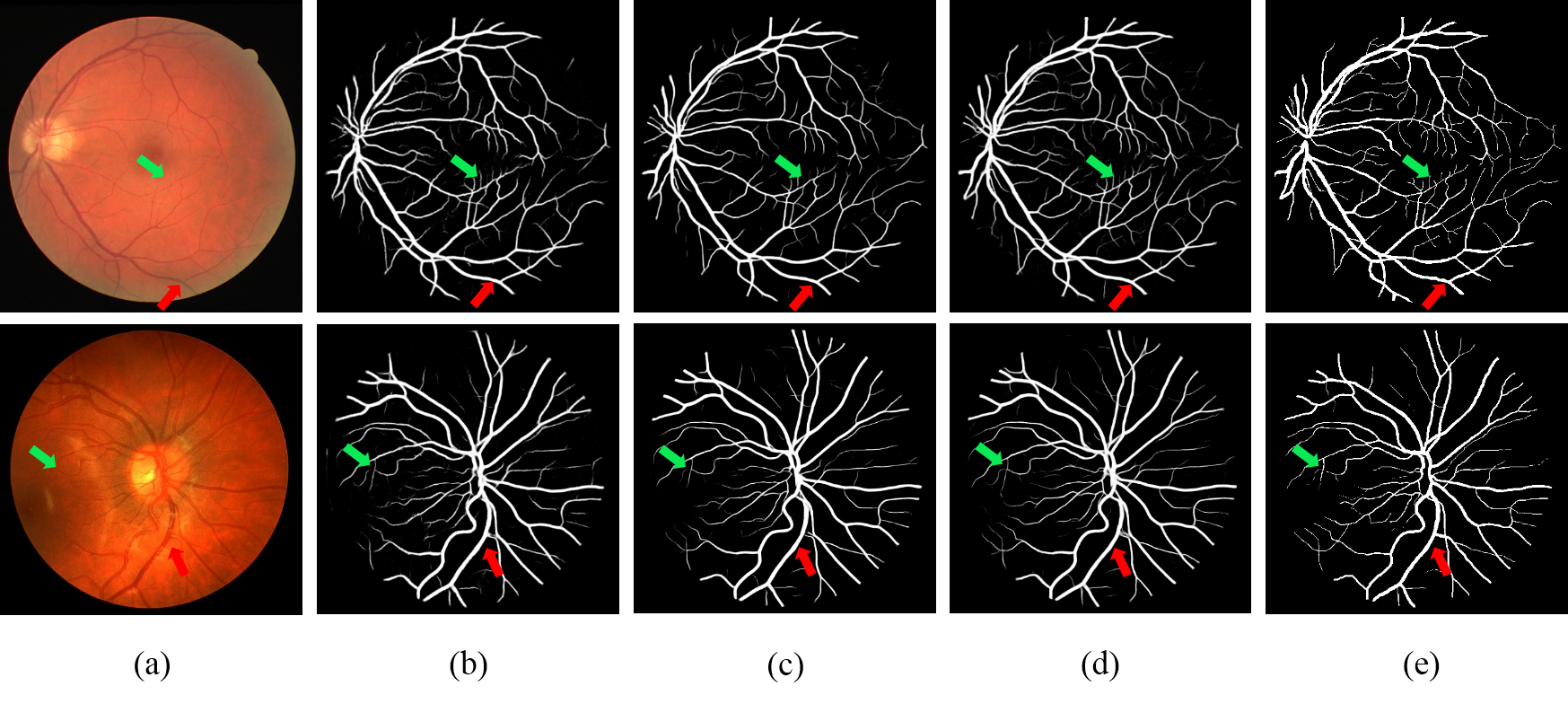}
\caption{Row 1 is for DRIVE dataset. Row 2 is for CHASE DB1 dataset. (a) Color fundus images, (b) segmentation results of Backbone, (c) segmentation results of Backbone+DropBlock, (d) segmentation results of RSAN, (e) corresponding ground truths.} \label{fig5}
\end{figure}

\begin{table}[htbp]
  \centering
  \caption{Experimental results on CHASE DB1.(*The results is obtain from \cite{14})}
    \begin{tabular*}{\hsize}{@{}@{\extracolsep{\fill}}cccccc@{}}
    \toprule
    Datasets & \multicolumn{5}{c}{CHASE DB1} \\
    \midrule
   Metrics & SEN & SPE & F1 & ACC & AUC \\
    \midrule
    U-Net \cite{5} * & 0.8288 & 0.9701 & 0.7783 & 0.9578 & 0.9772 \\
    Backbone & 0.7843 & \textbf{0.9844} & 0.7781 & 0.9718 & 0.9805 \\
    Backbone+DropBlock & 0.8330 & 0.9830 & 0.7990 & 0.9736 & 0.9871 \\
    \textbf{RSAN} &\textbf{0.8486} & 0.9836 & \textbf{0.8111} & \textbf{0.9751} & \textbf{0.9894} \\
    \bottomrule
    \end{tabular*}%
  \label{tab:addlabe2}%
\end{table}%

Finally, we compare our proposed RSAN with several existing state-of-the-art methods. We summarize the release year of the different methods and their performance on DRIVE and CHASE DB1, as shown in Tables 3 and 4, respectively. From the results in these tables, our proposed RSAN achieves the best performance on both datasets. In detail, on the DRIVE and CHASE DB1, our RSAN has the highest AUC (0.34$\%$/0.34$\%$ higher than the best before), the highest accuracy (1.13$\%$/0.90$\%$ higher than the best before) and the highest sensitivity. Besides, F1 and specificity are comparable in general. The above results mean that our method achieves the state-of-the-art performance for retinal vessel segmentation.
\begin{table}[htbp]
  \centering
  \caption{Results of RSAN and other methods on DRIVE dataset}
    \begin{tabular*}{\hsize}{@{}@{\extracolsep{\fill}}ccccccc@{}}
    \toprule
    Methods & Year & SEN & SPE & F1 & ACC & AUC \\
    \midrule
    Orlando et. al. \cite{12} & 2017  & 0.7897 & 0.9684 & 0.7857 & 0.9454 & 0.9506 \\
    Yan et al. \cite{13} & 2018  & 0.7653 & 0.9818 & N.A & 0.9542 & 0.9752 \\
    R2U-Net \cite{14} & 2018  & 0.7799 & 0.9813 & 0.8171 & 0.9556 & 0.9784 \\
    LadderNet \cite{2} & 2018  & 0.7856 & 0.9810 & 0.8202 & 0.9561 & 0.9793 \\
    MS-NFN \cite{3} & 2018  & 0.7844 & 0.9819 & N.A & 0.9567 & 0.9807 \\
    DEU-Net \cite{4} & 2019  & 0.7940 & 0.9816 & \textbf{0.8270} & 0.9567 & 0.9772 \\
    Vessel-Net \cite{15} & 2019  & 0.8038 & 0.9802 & N.A & 0.9578 & 0.9821 \\
    \textbf{RSAN} & \textbf{2020} & \textbf{0.8149} & \textbf{0.9839} & 0.8222 & \textbf{0.9691} & \textbf{0.9855} \\
    \bottomrule
    \end{tabular*}%
  \label{tab:addlabe3}%
\end{table}%

\begin{table}[htbp]
  \centering
  \caption{Results of RSAN and other methods on CHASE DB1 dataset.}
    \begin{tabular*}{\hsize}{@{}@{\extracolsep{\fill}}ccccccc@{}}
    \toprule
    Methods & Year & SEN & SPE & F1 & ACC & AUC \\
    \midrule
    Orlando et. al. \cite{12} & 2017  & 0.7277 & 0.9712 & 0.7332 & 0.9458 & 0.9524 \\
    Yan et al. \cite{13} & 2018  & 0.7633 & 0.9809 & N.A &  0.9610 & 0.9781 \\
    R2U-Net \cite{14} & 2018  & 0.7756 & 0.9820 & 0.7928 & 0.9634 & 0.9815 \\
    LadderNet \cite{2} & 2018  & 0.7978 & 0.9818 & 0.8031 & 0.9656 & 0.9839 \\
    MS-NFN \cite{3} & 2018  & 0.7538 & \textbf{0.9847} & N.A &  0.9637 & 0.9825 \\
    DEU-Net \cite{4} & 2019  & 0.8074 & 0.9821 & 0.8037 & 0.9661 & 0.9812 \\
    Vessel-Net \cite{15} & 2019  & 0.8132 & 0.9814 & N.A & 0.9661 & 0.9860 \\
    \textbf{RSAN} & \textbf{2020} & \textbf{0.8486} & 0.9836 & \textbf{0.8111} & \textbf{0.9751} & \textbf{0.9894} \\
    \bottomrule
    \end{tabular*}%
  \label{tab:addlabel4}%
\end{table}%

\section{Discussion and Conclusion}
Residual Spatial Attention Network (RSAN) is presented in this paper to be utilized to accurately segment retinal blood vessel in fundus images. RSAN exploits the pre-activation residual block integrated with DropBlock, which can effectively extract more complex vascular features and prevent overfitting. In addition, the newly designed Residual Spatial Attention Block (RSAB) significantly improves the network's representation capability via introducing the spatial attention mechanism. We evaluate the RSAN on two datasets, including DRIVE and CHASE DB1, and the results indicate that RSAN reaches the state-of-the-art performance. The improvement of RSAN's performance in retinal blood vessel segmentation, especially the improvement of AUC and ACC indicators, is of great significance in the clinical detection of early eye-related diseases. It is worth mentioning that RSAN has not yet considered the connectivity of blood vessels, which is also the focus of our next work.

\end{document}